\documentclass[preprint]{aastex}

\slugcomment{Accepted by ApJ on 8 October 2010}

\shorttitle{Infrared Molecular Hydrogen  Emission from Protostars}
\shortauthors{Greene, Barsony, \& Weintraub}

\begin{document}

\title{Near-IR H$_{2}$ Emission of Protostars: Probing Circumstellar
Environments\footnote{The data presented herein were obtained
at the W.M. Keck Observatory from telescope time allocated to the
National Aeronautics and Space Administration through the agency's
scientific partnership with the California Institute of Technology and
the University of California.  The Observatory was made possible by
the generous financial support of the W.M. Keck Foundation.}}

\author{Thomas P. Greene}
\affil{NASA Ames Research Center, M.S. 245-6, Moffett
Field, CA 94035}
\email{tom.greene@nasa.gov}

\author{Mary Barsony\altaffilmark{2}}
\affil{Department of Physics and Astronomy, San Francisco
State University, 1600 Holloway Drive, San Francisco, CA 94132}
\email{mbarsony@SpaceScience.org}

\and

\author{David A. Weintraub}
\affil{Department of Physics and Astronomy, Vanderbilt University, Nashville, TN 37235}
\email{david.a.weintraub@vanderbilt.edu}

\altaffiltext{2}{Space Science Institute, 4750 Walnut Street, Suite 205, Boulder, CO 80301}

\begin{abstract}

We present new observations of near-infrared molecular hydrogen
(H$_{2}$) line emission in a sample of 18 Class I and flat-spectrum low
mass protostars, primarily in the Tau-Aur and $\rho$ Oph dark clouds.
The line emission is extended by up to several arcseconds (several
hundred AU) for most objects, and there is little night-to-night
variation in line strength coincident with the continuum point source.
Flux ratios of H$_{2}$ $v = 2-1$ $S(1)$ and $v = 1-0$ $S(1)$ lines are
consistent with this emission arising in jets or winds in many objects.
However, most objects have only small offsets (under 10 km s$^{-1}$)
between their H$_{2}$ and photospheric radial velocities. No objects
have line ratios which are clearly caused solely by UV excitation, but
the H$_{2}$ emission of several objects may be caused by UV or X-ray
excitation in the presence of circumstellar dust. There are several
objects in the sample whose observed velocities {\it and} line fluxes
suggest quiescent, non-mechanical origins for their molecular hydrogen
emissions. Overall we find the H$_{2}$ emission properties of these
protostars to be similar to the T Tauri stars studied in previous
surveys.

\end{abstract}


\section{Introduction}

Embedded low-mass protostars have been identified from their infrared
(IR) energy distributions for over two decades, but their high
extinctions and relatively small sizes make it very difficult to
observe how radiation from the central protostars interacts with their
inner, pre-planetary circumstellar disks. Recent high sensitivity, high
resolution spectroscopic surveys have revealed the detailed {\it
stellar} properties of significant numbers of these objects
\citep{DGCL05,WH04,GL02}, but less is known about the physical
conditions in their disks, especially of the gaseous component, or
about their inner disks where planets might form and migrate. Here, we
explore the gaseous component of the innermost regions of self-embedded
(Class I and flat-spectrum; hereafter FS) protostars via
high-resolution near-infrared (NIR) spectroscopy of the most abundant
molecule in proto-planetary disks, molecular hydrogen (H$_2$).
Near-infrared H$_2$ transitions are used, since the high extinctions to
these sources preclude UV spectroscopy.

The NIR ro-vibrational lines of H$_{2}$ are good tracers of
physical conditions in inner circumstellar disks and winds close to
protostars.  There are several different mechanisms that might be responsible
for the production of NIR H$_2$ emission lines in late-stage protostars.
These include: {\it i)} shock heating of the ambient medium by winds or jets,
{\it ii)} X-ray heating, or {\it iii)} UV-heating, level
pumping and fluorescence.


Protostars must accrete mass at mean rates of $\sim10^{-6} - 10^{-5}
M_{\sun}$ yr$^{-1}$ to assemble themselves on time scales of several
$10^{5}$ yr, and these high rates produce significant UV flux when the
accreting matter impacts the stellar surface \citep[e.g.,][]{GCMH00}.
Protostars and T Tauri stars are also known to be strong and variable
sources of X-ray emission \citep[e.g.,][] {IKT01, INTKT03, GBAAB07,
FSSMPT09}. Both UV and X-rays can excite vibrational states of 
H$_{2}$, producing NIR line emission \citep[e.g.,][]{GD95,
NATNM07}. Many protostars and T Tauri stars also shed mass in
jets that drive molecular outflows, and these jets also frequently
excite NIR vibrational H$_{2}$ line emission \citep{ZMR98}.

The most commonly encountered excitation mechanism for NIR H$_2$
emission associated with protostars to date has been
shock-excitation\footnote{See {\tt
http://www.jach.hawaii.edu/UKIRT/MHCat/} for an up-to-date listing of
``Molecular Hydrogen Emission-Line Objects (MHOs) in Outflows from
Young Stars.''} \citep[e.g.,][]{DAV10}, 
consistent with many self-embedded
protostars being associated with large-scale molecular outflows
\citep[e.g.,][]{MSWTK92}.  These outflows are understood to be driven
by powerful stellar winds \citep[]{MC93}.  Such winds
are often detected as blue-shifted absorption components in forbidden emission
lines or in the HeI 1.0830 $\mu$m line associated with protostars 
and young T Tauri stars \citep[e.g.,][]{ERM93,KWA07}. 
Winds are also believed to cause the shocked IR H$_{2}$
emission seen in jets from heavily extinguished young stellar objects.
Such stellar winds are inferred to be driven by mass
accretion onto young stars.  
Jets detected in NIR H$_2$ lines frequently are displaced in space and in radial
velocity from photospheric absorption features and usually have relatively large
linewidths of a couple dozen km s$^{-1}$ or more. For example,
in a recent VLT/ISAAC pilot study of the H$_2$ 1$-$0
S(1) emission from embedded Class I sources, 
mean intensity-weighted velocities were blue-shifted by -90 to -10 km s$^{-1}$
and velocity widths of the lines varied from 
$\sim$ 45 km s$^{-1}$ to $\sim$ 80 km s$^{-1}$ \citep[]{CHR08}.

X-rays and UV may also excite molecular hydrogen emission in the inner
circumstellar environments of protostars. Both of these processes offer
a radiative alternative to the mechanical excitation excitation of the
NIR H$_2$ emission. If dominant, these excitation mechanisms may
produce emission lines with lower velocity widths or offsets than if
the H$_2$ were mechanically excited in shocks or jets. We now consider
the mechanisms, observational evidence, and implications for these
processes in protostellar environments.

It is now well established that Class I/FS protostars are copious X-ray
emitters (10$^{29}\le L_x\le 10^{31}$ erg s$^{-1}$, $\sim$100$-$1000
times more X-ray luminous than main-sequence stars), and emit higher
energy X-rays (4-6 keV vs. 1-2 keV) than their older, T Tauri cousins
\citep[e.g.,][]{CMFA95,GMFAC97,IKT01,FTGS07}. One would, therefore,
expect significant X-ray heating of the disk/inner envelope material
surrounding these central objects \citep[e.g.,][]{MPG09,MGN07}. X-rays
from young stellar objects (YSOs) can penetrate disk atmospheres to
fairly large surface densities and can ionize circumstellar gas at a
level greater than Galactic cosmic rays out to large distances
\citep[$\sim$10$^4$ AU;][]{GFMW05}.

The mechanism of H$_2$ excitation by X-rays requires impacts from
energetic electrons. X-rays impact hydrogen molecules, ejecting
electrons. These high energy electrons subsequently collide with and
ionize or dissociate ambient gas, losing kinetic energy in the process.
Some of electrons will eventually have energies appropriate to excite
ambient H$_2$ molecules into excited states instead of dissociating
them completely \citep[e.g.,][]{GD95, TLGD97, MHT96}. Direct evidence
for X-ray heating of gas other than H$_2$ in disks has been found for
several Class I sources in which the 6.4 keV line from neutral iron has
been detected \citep[and references therein]{GFPFMS07}.


Glassgold and co-workers especially have emphasized the importance of
X-ray heating of disk atmospheres out to relatively large distances
from the central source \citep{GNI07,GNI04}. Such heating can extend to
large distances because of disk flaring, first proposed to account for
the spectral energy distributions (SEDs) produced by dust
\citep{KH87,CG97}. More recent disk modeling, especially studies
involving disk chemistry, routinely use vertically stratified models,
with different gas and dust scale heights
\citep{DCHLC99,AZDH02,GH08,LHM10}. In X-ray heated disk models, there
is a low column density (N$_H\sim$10$^{20}$ cm$^{-2}$) surface layer of
hot (T$\sim$4000 K) gas that extends to $\gtrsim$10 AU radius \citep{NAJ09}.

 %
%
%
%
%
%

In the context of disk heating by X-rays, it must be noted that current
disk models assume irradiation by a central source with X-ray spectra
typical of T Tauri stars: with plasma having $kT_X=$1 keV and a
low-energy cut-off of 100 eV \citep[e.g.,][]{GNI07}. However, Class I
and FS sources are known to have harder X-ray spectra, with 4 keV $\le
kT_x \le$6 keV being typical \citep[e.g.,][]{IKT01}. Unsurprisingly,
the column densities of hydrogen gas inferred towards Class I/FS
sources from X-ray observations are $\sim\ N_H\ =\ 1-5 \times 10^{22}$
cm$^{-2}$, 1-2 orders of magnitude greater than inferred for T Tauri
stars. Finally, the quantity, L$_x$/L$_{bol}$, is systematically
smaller for Class I/FS objects relative to T Tauri stars, consistent
with the interpretation of higher accretion rates in these systems. The
higher accretion rates create higher optical depths in the X-ray
absorbing gas, obscuring the lower energy X-rays and producing
relatively lower L$_x$/L$_{bol}$ ratios. Higher accretion rates would
also lead to higher heating rates of the disk gas in self-embedded
protostars than in classical T Tauri star (CTTS) disks
\citep[e.g.,][]{DAL04}.

%
%
%
%
%

The possibility of UV excitation of the NIR H$_2$ lines also exists for
Class I/FS protostars, given their consistently higher accretion rates
relative to T Tauri stars. The excess UV continuum emission
observed in CTTSs has been modeled as being produced by the
impact of accretion columns onto the pre-main-sequence stellar surface
\citep[e.g.,][]{GCMH00}. This UV continuum excess could potentially
excite molecular hydrogen, producing emission lines in the
near-infrared. Strong Lyman-$\alpha$ emission from the central object
can irradiate the disk's surface, and, if H$_2$ is present at $\sim$
2000K, can excite the H$_2$ into electronic states which produce a rich
UV emission-line spectrum as observed in some T Tauri stars
\citep[]{HLWGJK06}. Finally, if the stellar EUV flux is sufficiently
strong, it can ionize hydrogen, which produces high temperatures
(T$\approx$ 10$^4$K) and small mean molecular weights at the disk
surface. Outside some critical radius, the gas becomes unbound, and a
slow, v$\approx$10 km s$^{-1}$, photoevaporative wind forms
\citep[]{ACP06a,ACP06b,GDH09,WOI09}.


Like molecular hydrogen, [NeII] emission can also arise from high
energy photons in dense circumstellar disks or outflows. [NeII] line
emission at 12.81 $\mu$m and [NeIII] emission at 15.5 $\mu$m was
predicted to be detectable in the case of X-ray heated disk gas. The
12.81 $\mu$m [NeII] line has subsequently been detected in a number of
young stars with the Spitzer Space Telescope \citep{GNI07,
PAS07,LAH07,FSSMPT09,NAJ10} and has been studied at high spectral
resolution from the ground in three objects \citep{HNHP07, NAJ09}. Although
a disk origin has been postulated or confirmed for the [NeII] emission observed in
the objects studied at high spectral resolution, there are also sources
in which [NeII] is detected in outflows close to the sources
\citep[e.g.,][]{NEU06,VAN09}. 

In general, gas motions in the inner circumstellar environments (within
$\sim$ 100 AU) of protostars have not been studied well except for a
very few cases of velocity resolved NIR CO observations, and this has
limited our understanding of how they accrete matter, form winds, shed
angular momentum, and disperse their natal circumstellar envelopes. We
seek to understand these processes as well as the UV and X-ray
radiation environments of their circumstellar disks by embarking on a
new study of protostars' NIR H$_{2}$ line strengths, morphologies, and
velocities.


H$_{2}$ NIR ro-vibrational line ratios yield excitation temperatures and also
provide clues to excitation mechanisms. 
The intensity ratio of $S(1)$ lines in the $v\ =\ 2\rightarrow1$ and
$v\ =\ 1\rightarrow0$ transitions is often used as a diagnostic. This
ratio has a value of 0.13 in shocked gas at 2000 K and a value of 0.54
for UV pumping. For X-ray excitation, this ratio is predicted to be
0.06 in gas of low fractional ionization (10$^{-4}$) and 0.54 in gas of
high fractional ionization (10$^{-2}$) \citep{GD95}. Any observed
variability in these emission lines can also provide clues to the
nature of their excitation and the location of the excited and emitting
gas.

Several recent studies have made good progress in diagnosing the nature
of H$_{2}$ excitation in the inner circumstellar disks of CTTSs using
observations of their NIR ro-vibrational emission lines. A growing body
of work shows that the NIR H$_{2}$ line radial velocities and
linewidths of many CTTS and Herbig AeBe stars are consistent with UV or
X-ray excitation in a circumstellar disk
\citep[]{WKB00,BWK02,BWK03,CAR07,BWSLK08}. By contrast, in a recently
completed NIR adaptive optics (AO) integral field spectroscopic study
of several CTTS, the H$_{2}$ emissions were most consistent with shocks
arising in winds \citep[]{BMTP08}.

Despite this progress in understanding H$_{2}$ emission in CTTSs,
little has been done to date in probing the nature of such emission in
more embedded Class I and flat--spectrum protostars. \citet{GL96}
reported that these protostars were significantly more likely to have
NIR H$_{2}$ emission than CTTSs, and \citet{DGCL05} found that 23 of 52
observed protostars showed NIR H$_2$ emission. The most embedded
protostars also have significant envelopes that may have enough column
density to generate observable H$_{2}$ emission if the hydrogen
molecules there receive sufficient radiative or mechanical energy. If
this envelope source existed, it would be a new source of emission not
present in CTTSs. Unfortunately, earlier studies did not have adequate
spectral resolution or spectral range to observe multiple NIR
ro--vibrational lines simultaneously with high spectral resolution and
good signal-to-noise: all necessary ingredients for measuring line
velocities and line ratios in order to diagnose excitation.

An early motivation for a new observational study was the identification 
of outflow drivers amongst late-stage protostars via detection of NIR H$_2$
emission at high spectral resolution observed directly towards the
putative powering source \citep{BAR05}. However, careful examination of
the H$_2$ $v = 1-0$ $S(0)$ line profiles (at 2.2235 $\mu$m) observed at
R$\sim$17,000 in the work of Doppmann et al. (2005), showed that a
substantial number of sources exhibited relatively narrow linewidths
($\Delta v\le$ 15 km s$^{-1}$). Furthermore, in many cases, the H$_2$
emission line centers were not significantly displaced from the central
object's radial velocity. Taken together, these two observations call
into question an outflow origin for the observed NIR H$_2$ emission in
some sources.

We have conducted a new study of Class I and flat-spectrum protostars 
with data sufficient for diagnosing the natures of their H$_{2}$ emission
in their inner circumstellar disks and envelopes. We report on the
sample and observations in \S 2 and present our analysis of these data
in \S 3. We discuss our results in \S 4 and summarize our work in \S 5.

\section{Observations and Data Reduction}

High resolution NIR spectra of 18 Class I and flat-spectrum protostars
previously searched for H$_{2}$ $v = 1-0$ $S(0)$ emission by
\citet{DGCL05} were re-observed with the Keck~II telescope on Mauna Kea,
Hawaii using its NIRSPEC multi-order cryogenic echelle facility
spectrograph \citep{McLeanetal98}. This included 17 objects found to
exhibit H$_{2}$ $v = 1-0$ $S(0)$ emission in this previous study, and one
(03260+311A) without. All new spectra were acquired on 2007 June 24 and
25 UT ($\rho$ Oph and Ser objects) and 2008 January 24 and 25 (Tau-Aur
and Per objects). The late type (K1 -- M2.5) dwarfs HD 20165, HD 28343,
and HD 285968 with precision velocities measured by \citet{NMBFV02} were
observed on 2008 January 24 UT to serve as radial velocity references.

Spectra were acquired with a 0\farcs58 (4-pixel) wide slit, providing
spectroscopic resolution $R \equiv \lambda / \delta \lambda$ = 18,000
(16.7~km~s$^{-1}$).  The plate scale was 0\farcs20 pixel$^{-1}$ along
the 12$\arcsec$ slit length, and the seeing was typically
0\farcs5--0\farcs6.  The NIRSPEC gratings were oriented to observe the
2.1218 $\mu$m H$_{2}$ $v = 1-0$ $S(1)$, 2.2477 $\mu$m H$_{2}$ $v = 2-1$
$S(1)$, and 2.386 $\mu$m H$_{2}$ $v = 3-2$ $S(1)$ lines on the
instrument's 1024 $\times$ 1024 pixel InSb detector array in a single
exposure.  The 2.2233 $\mu$m H$_{2}$ $v = 1-0$ $S(0)$ line previously
observed by \citet{DGCL05} was also captured in this grating setting.
The NIRSPEC-7 blocking filter was used to image these orders on the
detector.  NIRSPEC was configured to acquire simultaneously multiple
cross-dispersed echelle orders 32--37 (2.05--2.40~$\mu$m,
non-continuous) for all objects.  Each order had an observed spectral
range $\Delta \lambda \simeq \lambda / 67$ ($\Delta v \simeq$ 4450
km~s$^{-1}$).

The slit was held physically stationary during the exposures and thus
rotated on the sky as the non-equatorially-mounted telescope tracked when
observing. Data were acquired in pairs of exposures of durations from
180--600~s each, with the telescope nodded 3$\arcsec$ or 6$\arcsec$ along
the slit between frames so that object spectra were acquired in all
exposures. Most of the targets were observed twice on consecutive nights.
The observation dates, total integration times, slit angles and
coordinates of all Class I and flat-spectrum objects are given in
Table~\ref{tbl-1}. The early-type (B9--A0) dwarfs HD 28354, HR 6070, HR
5993, and HD 168966 were observed for telluric correction of the target
spectra. The telescope was automatically guided with frequent images from
the NIRSPEC internal ``SCAM'' IR camera during all exposures of more than
several seconds duration. Spectra of the internal NIRSPEC continuum lamp
were taken for flat fields, and exposures of the Ar, Ne, Kr, and Xe lamps
were used for wavelength calibrations.

All data were reduced with IRAF. First, object and sky frames were
differenced and then divided by normalized flat fields.  Next, bad
pixels were fixed via interpolation, and spectra were extracted with
the APALL task.  Spectra were wavelength calibrated using low-order
fits to lines in the arc lamp exposures, and spectra at each slit
position of each object were co-added.  Instrumental and atmospheric
features were removed by dividing wavelength-calibrated object spectra
by spectra of early-type stars observed at similar airmass at each
slit position.  Final spectra were produced for each night by
combining the spectra of both slit positions for each object and then
multiplying them by spectra of 10,000 K blackbodies to rectify the
spectral shapes induced when dividing by the telluric stars with that
effective temperature. Spectra acquired on different nights were not
combined in any way.

\section{Analysis and Results}

The spectra of all objects in Table~\ref{tbl-1} were analyzed by
measuring their H$_{2}$ $v = 1-0$ $S(1)$, $v = 2-1$ $S(1)$, and $v = 1-0$
$S(0)$ line fluxes, H$_{2}$ line radial velocities, and the radial
velocities of photospheric absorption lines. The 2.386 $\mu$m H$_{2}$
$v = 3-2$ $S(1)$ line is in a region of poor atmospheric transmission and
was not significantly detected in any object. The values of the H$_{2}$
line properties and their night-to-night variations were analyzed for
physical insights into their excitation as described in this section.

\begin{figure}
\includegraphics[ scale=0.8,angle=0]{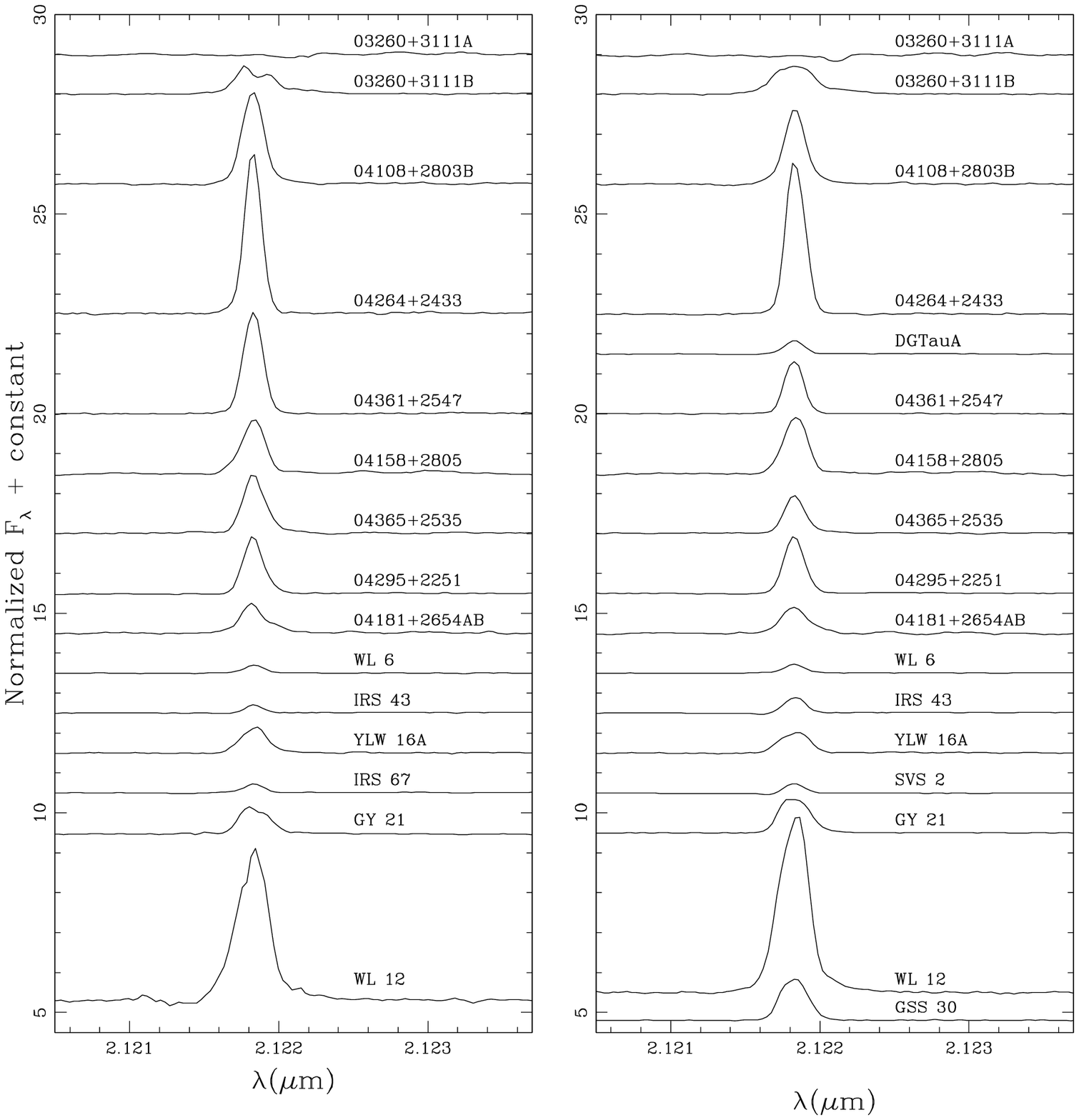}
\caption{H$_{2}$ $v = 1-0$ $S(1)$ line spectra. Spectra acquired on
2007 Jun 24 and 2008 Jan 24 (first epoch) appear in the left panel, and spectra
acquired on 2007 Jun 25 and 2008 Jan 25 (second epoch) appear in the right panel.
These spectra were extracted only in the region containing the mostly 
point source continuum emission; any extended H$_{2}$ emission is not
included.
\label{fig1}}
\end{figure}

Spectra of the H$_{2}$ $v = 1-0$ $S(1)$ region are shown for all objects
in Figure~\ref{fig1}; the first epoch (2007 Jun 24 and 2008 Jan 24) is
shown in the left panel and the second (2007 Jun 25 and 2008 Jan 25) in
the right. Of the 18 objects, 14 were observed in both epochs. When
present, the $v = 1-0$ $S(1)$ line was the strongest H$_{2}$ feature
observed in all objects. All emission spectra shown in the figures and
all derived strength and velocity values presented in Table~\ref{tbl-2}
are for the H$_{2}$ emission that was spatially coincident with each
object's continuum source, typically limited by the $0\farcs6$ seeing,
corresponding to a spatial extent of $\sim 100$ AU for most sources.

\subsection{H$_{2}$ Emission Morphologies}

The spatial extent of the H$_{2}$ emission along the spectrograph slit
was measured for each object by examining the differenced and
flat-fielded spectral images that were created prior to spectral
extraction.  The non-rotating spectrograph slit was projected onto the
sky at different position angles each night, so the two
observational epochs sample the spatial extent of any extended
emission differently (see position angles in Table~\ref{tbl-1}).

The angular extent of the H$_{2}$ emission along the length of the slit
is reported in Table~\ref{tbl-2}) for each observation. Ten of the 18
protostars showed H$_{2}$ emission that was over $1\arcsec$ in spatial
extent in at least one observation (at least one position angle). Of the
14 objects observed in two epochs, 6 had H$_{2}$ $v = 1-0$ $S(1)$
emission extended by $\sim 1\arcsec$ or less on both nights. Therefore
the molecular hydrogen emission of these 6 objects is contained within
about 70 AU ($\sim 0\farcs5$) of their central stars. The highest H$_{2}$
surface brightness was generally found to be coincident with each
object's unresolved point source, but in some cases the extended emission
may have more integrated flux and luminosity than the point source
component. \citet{BMTP08} found this to be true in their AO
integral field study of several CTTSs that sampled their immediate
circumstellar environments completely instead of evaluating just two slit
position angles as done in this study.

Objects with significantly different spatial extents between epochs
(position angles) may have their H$_{2}$ emission confined to
asymmetrical structures like jets. 03260+3111B, 04158+2805, and
04264+2433 are examples of objects with these potentially jet-like 
morphologies. 03260+3111A shows significantly extended emission, but
this is spatially displaced from the stellar continuum which has no
coincident H$_{2}$ emission.


\subsection{Variability}

We found H$_{2}$ emission in all of the 17 objects also found to have
H$_{2}$ emission at earlier epochs by \citet{DGCL05}. The one
source without emission (03260+3111A) in that earlier study also did not
show H$_{2}$ emission in our new observations. This indicates that
these these protostars do not terminate or initiate their H$_{2}$
emissions on time scales as short as 5 -- 10 years, suggesting that
their H$_{2}$ lines are emitted over physically large regions or else 
are excited by processes with relatively stable fluxes.

Fourteen of the 18 objects were observed twice, with the two
observations spaced approximately 24 hr apart.  The position angles of
both observations were sometimes similar (within $\sim 10 \deg$), but
often they varied by $60\deg$ or more (Table ~\ref{tbl-1}).  Therefore
the different slit position angles must be considered when comparing
H$_{2}$ line fluxes measured on the different dates to determine
whether any differences are due to true temporal variation or the
rotation of different spatial features (i.e., jets) onto or off of the 
slit. However, the measured equivalent H$_{2}$ emission widths and
line velocities were similar for both observations of each object (see
Table ~\ref{tbl-2}), so there appears to be little temporal variation
on $\sim1$ day time scales for the emission that is spatially confined
to be coincident with the protostar continuum source confined within
the slit width.

\subsection{Radial Velocities and Line Widths}

Stellar photospheric radial velocities were computed from the
continuum absorption lines of all observed objects.  First, the object
spectra were cross-correlated (using fxcor in IRAF) with spectra of
the radial velocity standards, always HD 28343 (K7 V) and sometimes HD
20165 (K1 V) or HD 285968 (M2.5 V), using up to 4 spectral orders that
contained photospheric lines but no emission lines.  Heliocentric
$V_{LSR}$ radial velocities were computed for each object by summing
the radial velocity shift measured with the cross-correlations, the
mean measured radial velocities of the 3 standards \citep{NMBFV02},
and the $V_{LSR}$ corrections for the objects and standards at their
time of observation (from the IRAF rvcorrect task).  The $V_{LSR}$
radial velocities computed in this way generally agreed well with
those of the 10 objects also reported by \citet{CGDL06}.  We observed
these 10 objects a total of 18 times, and we measured the offset from
the \citet{CGDL06} velocities to be $2.2 \pm 3.5$ km s$^{-1}$.  This
offset is consistent with zero, and the standard deviation is not
surprising given the 17~km~s$^{-1}$ spectral resolution and the
generally heavily veiled spectra.

\begin{figure} 
	\includegraphics[
scale=0.8,angle=0]{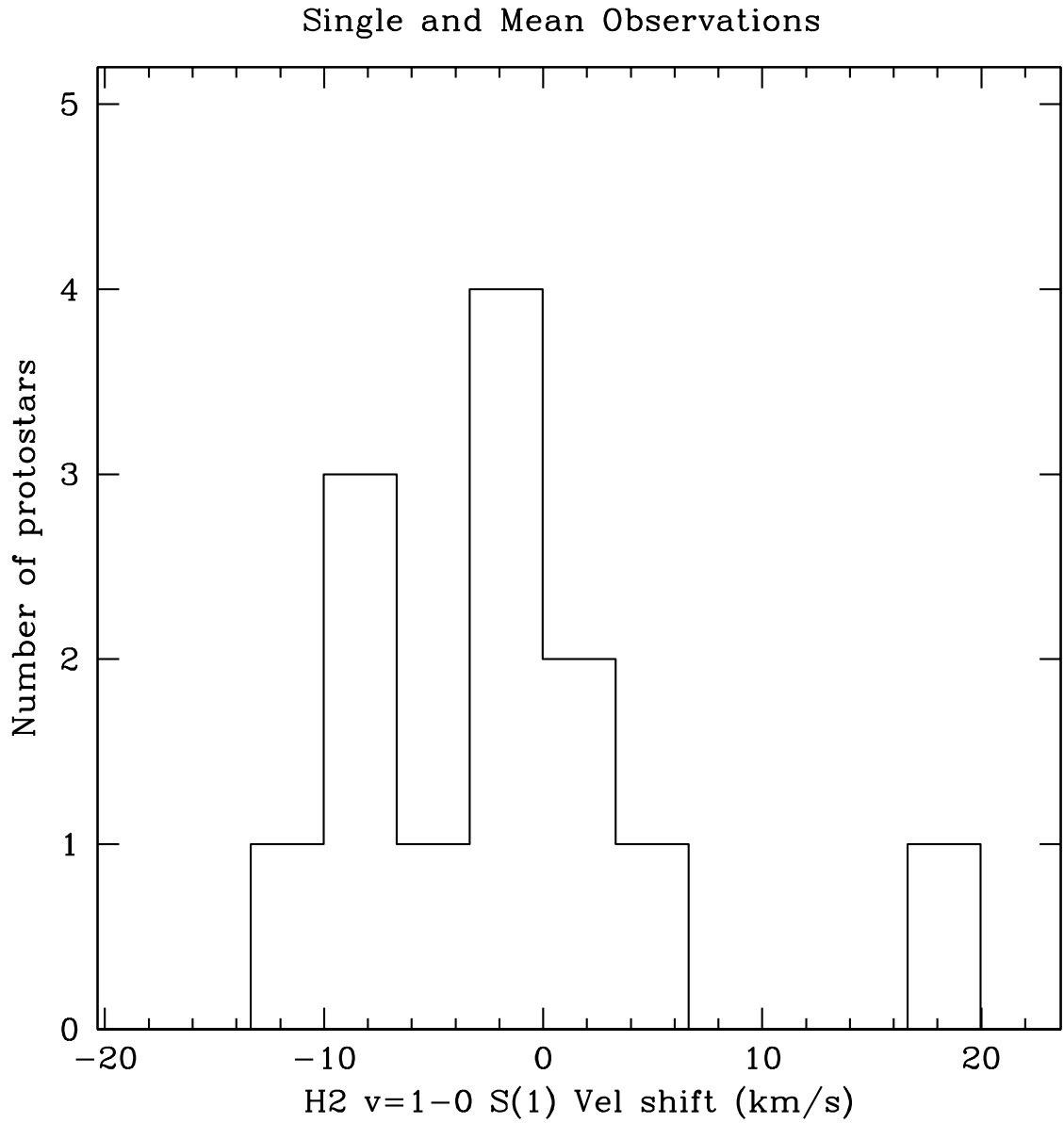} \caption{Histogram
of mean H$_{2}$ $v = 1-0$ $S(1)$ line velocities minus photospheric
velocities for all observations.\label{fig2}} \end{figure}

Radial velocities of H$_{2}$ lines were computed by measuring the
central wavelengths of Gaussian profiles fit to the lines using the
IRAF splot task and then converting these values to velocities.  These
observed radial velocities were converted to $V_{LSR}$ values by adding
the $V_{LSR}$ correction offset computed for each object as described
above.  The velocities of the H$_{2}$ lines relative to photospheric
lines of each object were computed by differencing these two $V_{LSR}$
radial velocities, and the results are shown in Table~\ref{tbl-2} and
Figure~\ref{fig2}.  FWHM velocity widths of the $v = 1-0$ $S(1)$ lines
were computed by subtracting the instrumental line width of 17 km
s$^{-1}$ in quadrature from the FWHM values of the Gaussian fits.
These resultant line widths are also reported in Table~\ref{tbl-2},
and their histogram is shown in Figure~\ref{fig3}.  Objects observed
twice on successive nights had similar radial and FWHM velocities
(Table~\ref{tbl-2}), so these values were averaged to reduce noise in
Figures~\ref{fig2} and \ref{fig3}.  The other H$_{2}$ lines generally
had similar FWHM values, but the $v = 1-0$ $S(1)$ lines had the highest
signal-to-noise, so only those values are presented.  We estimate that
all reported velocities have uncertainties of a few km s$^{-1}$.

\subsection{H$_{2}$ Line Fluxes and Ratios}

Line luminosities were estimated by scaling the relative fluxes
measured in each H$_{2}$ line to the spatially coincident 2.2 $\mu$m
continuum and multiplying this by the absolute 2.2 $\mu$m continuum
flux estimated from each object's 2MASS K-band magnitude after
correcting for extinction. Extinctions were calculated by de-reddening
each object's $JHK$ 2MASS magnitudes to the CTTS locus \citep{MCH97}.
Extinctions were estimated at each H$_{2}$ line wavelength using $A_{v}
= 9.09 [(J-H) - (J-H)_{0}]$, $A_{k} = 0.09 A_{v}$, and $A_{\lambda}
\propto \lambda^{-1.9}$. These values were computed by and derived from
\citet{CPEF81} for the CIT photometric system, which is essentially
identical to that of 2MASS \citep{C01}.

Distances were assumed to be 140 pc for Tau-Aur \citep{KDH94}, 140 pc
for $\rho$ Oph \citep{M08}, 260 pc for SVS 2 in Serpens \citep{SCB96},
and 320 pc for the 03260+3111 objects in Perseus \citep{H98}.  H$_{2}$
line equivalent widths and luminosities are presented in
Table~\ref{tbl-2}, and a histogram of luminosities of the H$_{2}$ $v =
1-0$ $S(1)$ line is shown in Figure ~\ref{fig4}.  A histogram of
H$_{2}$ $1-0 / 2-1$ $S(1)$ line ratios is presented in Figure
~\ref{fig5}.  As done in previous figures, values derived from
observations of objects acquired on successive nights are averaged in
these figures as well.  Note that the H$_{2}$ line ratio value plotted
in Figure ~\ref{fig5} is the inverse of the values presented in
Table~\ref{tbl-2} (column 8).

\subsection{Correlations}

We examined whether correlations exist between measured H$_{2}$ line
properties and other protostellar activity indicators in an attempt to
isolate the origins of the H$_{2}$ line emissions. 

First we computed the correlations between the H$_{2}$ line properties
measured in this new survey.  The FWHM velocity widths and the $1-0 /
2-1$ $S(1)$ line ratios in Table~\ref{tbl-2} have a correlation
coefficient of 0.61, which improves to 0.71 when both observations of
a single object are averaged into single points.  This value drops to
only 0.17 (individual or mean values) if the observations of
03260+3111B are excluded; its high FWHM velocity and relatively high
$1-0 / 2-1$ $S(1)$ line ratio drives this correlation.  Thus the
object sample as a whole does not show a good correlation between its
H$_{2}$ FWHM velocities and $1-0 / 2-1$ $S(1)$ line ratios. We then
examined correlations between H$_{2}$ FWHM velocities and velocity
offsets between H$_{2}$ and photospheric lines. These values had
correlation coefficients of -0.43 and -0.47 for individual and
averaged values, respectively. The square of the correlation
coefficient is below 0.25 in both cases, indicating that less than
25\% of the variance of the two quantities are in common for this
sample of protostars, a poor correlation. 

Next we correlated the protostars' H$_{2}$ line properties with other
physical characteristics measured in other studies. Twelve of the 18
objects have measured X-ray luminosities or upper limits
\citep{GBAAB07, FSSMPT09}, and the correlation coefficient between the
logarithms of their H$_{2}$ $1-0$ $S(1)$ line luminosities and the
logarithms of their X-ray luminosities is -0.20, suggesting a very weak
or nonexistent inverse correlation between these properties. Finally,
we re-analyzed the spectra of \citet{DGCL05} and used their equivalent
width measurements of H$_{2}$ $1-0$ $S(0)$ and HI Br $\gamma$ emission
lines to evaluate the correlation of these properties in that somewhat
earlier epoch. We found that these values had a correlation coefficient
of 0.20, indicating another poor correlation. In summary, we find
little correlation among the NIR H$_{2}$ line emission properties or
between these properties and other young stellar activity indicators.


\section{Discussion}

The different radiative and collisional excitation mechanisms of
H$_{2}$ are well matched to the radiative and mass flux processes in
the environments of protostars and T Tauri stars.  We now interpret
the results of the preceding analysis in terms of several of these
possible processes in order to constrain the H$_{2}$ excitation
mechanisms of the sample and to understand better the circumstellar
environments of these protostars.

\subsection{Emission Morphologies, Variability, and Velocities}

In addition to their compact H$_{2}$ line emissions, all but 4 of the
18 objects also showed $v = 1-0$ $S(1)$ molecular hydrogen emission
extended by $\sim 1\arcsec$ or more along the slit (see
Table~\ref{tbl-2}), corresponding to 70 AU ($\sim 0\farcs5$) or more
projected radial distance. It is unlikely that UV radiation could
travel that far from the central protostars without significant
attenuation by gaseous and dusty envelopes, so it is likely that this
extended emission is excited by either stellar winds or high energy
X-rays. The expected small size of the UV emission region on the
protostellar photosphere also suggests that UV may not be a good
candidate for exciting the observed steady and long-lived H$_{2}$ line
emissions. Two objects, GY 21 and IRS 43, have extended $1-0$ $S(1)$
emission with broad line widths, FWHM $\gtrsim$ 40 km s$^{-1}$, about
twice that of the compact emission spatially coincident with their
stellar continua. The extended emissions of these 2 objects are good
candidates for excitation in shocks caused by stellar winds.

The relatively stable values of the point source H$_{2}$ emission over
1 day and several-year time scales (see \S 3.2 and Table~\ref{tbl-1})
also provide clues to the nature of these emissions. Numerous Class I
and FS protostars have been observed to undergo rapid (several hr),
large amplitude X-ray emission variations \citep[e.g.,][] {IKT01,
INTKT03, GBAAB07, FSSMPT09}. This X-ray flaring of several protostars
(e.g., IRS 43) has also been observed to appear or disappear in data
taken $\sim$ 5 -- 10 years apart. If these X-ray flares were exciting
H$_{2}$ close to the stars, then it is likely that we would see
significant night-to-night or year-to year variations in their
point-source near-IR line fluxes, but this is not seen in our data. The
very stable observed molecular hydrogen emission is more consistent
with mechanical (wind or jet) excitation as well as excitation by
steady, non-flaring X-ray emission from the protostars.

The preceding analysis of the velocity widths and velocity shifts of
the H$_{2}$ line emission in \S 3.3 also provides clues to the nature of its
excitation.  Collisional excitations in jets or winds are likely to
result in H$_{2}$ emission line radial velocities displaced from
photospheric absorption lines by over 10 km s$^{-1}$, exhibition of
broad line wings, or large full-width half maximum velocities of
several 10 km s$^{-1}$ or more \citep{MHT96, MGTK00, NATNM07, BMTP08}.  Jets
are also often significantly collimated and spatially extended, making
them easy to identify in one or two dimensional spectral images
\citep[e.g.,][]{SG03}.

The significant spatial extension of the H$_{2}$ line emission of many
objects is consistent with collisional excitation in winds or by jets.
However, only 5 of 13 protostars show radial velocity offsets between
their H$_{2}$ lines and photospheric absorption lines with absolute
value of greater than 4 km s$^{-1}$ (see Fig.~\ref{fig2}). This value
is similar to the 3.5 km s$^{-1}$ uncertainty we measured for radial
velocity standards (see \S 3.3), so we do not consider velocity offsets
less than 4 km s$^{-1}$ to be significant. Only 3 protostars have
velocity offsets of at least 10~km~s$^{-1}$, significant at about the
3-$\sigma$ confidence level or greater. This evidence suggests that
collisional excitation in jets is unlikely to be the molecular hydrogen
excitation mechanism in most objects (except the 3 with significant
velocity differences).


However, the on-source H$_{2}$ $v = 1-0$ $S(1)$ line FWHM line widths
of all 17 protostars with this on-source feature are broader than 10 km
s$^{-1}$ (see Fig.~\ref{fig3}). Six of the emitting objects (35\%) exhibit
FWHM greater that 20 km s$^{-1}$. All 6 of these objects also show
spatially extended H$_{2}$ emission many tens of AU from their central
stars; this combination of factors makes them good candidates for
collisional excitation in jets or winds. Seven of the emitting objects
(41\%) have FWHM line widths below 16 km s$^{-1}$. This is similar to
the 9 to 14 km s$^{-1}$ line widths found by \citet{BWK03} and
\citet{BWSLK08} for 8 of 10 T Tauri stars found to have H$_{2}$
emission, which they interpreted as evidence for quiescent emission in
circumstellar disks. We conclude that the molecular hydrogen emission
line velocities and FWHM values of at least 3 to 6 of the 17 emitting
objects are consistent with collisional excitation in jets, and at
least 7 or 8 objects have H$_{2}$ velocity parameters consistent with
quiescent (non-collisional) excitation.

\subsection{Emission Line Strengths and Ratios}

NIR vibrational H$_{2}$ line ratios are also sensitive to the gas
excitation levels and excitation mechanisms.  \citet{GD95} show that
the ratios of the H$_{2}$ $v = 1-0$ $S(1)$ to $v = 2-1$ $S(1)$ lines are
relatively sensitive to excitation mechanisms.  They compute the
ratios of these 2 lines to be 1.9 for UV excitation, 7.7 for shocked
gas at $T=2000$ K, and 16.7 for X-ray excitation of low ionization
H$_{2}$.  \citet{BD87} also found similar differences between UV and
shock excitation of H$_{2}$.  However, there are limits to the
usefulness of these ratios as diagnostics of excitation in
circumstellar disks.  In practice it is difficult to distinguish
between shocked and X-ray excited H$_{2}$ emission from examining only
a few NIR lines.  Collisions will thermalize the excitation levels
of H$_{2}$ in a sufficiently dense gas, so $v = 1-0$ $S(1)$ to $v = 2-1
S(1)$ line ratios indicative of cold-to-warm gas in equilibrium are
not always useful for distinguishing between excitation mechanisms
\citep{GD95, MHT96, TLGD97, NATNM07, BMTP08}.

However, these line ratios may be useful for diagnosing excitation
processes in extreme cases. \citet{BD87} note that fluorescent UV
excitation of H$_{2}$ produces significant population of vibrational
levels $v \geq 2$ and therefore strong emission in the $v = 2-1$ $S(1)$
line when not thermalized in a high density environment. This level
population can be characterized by a temperature of $T_{\rm vib} \approx
6000 - 9000$ K, much larger than the $T_{\rm vib} = T \simeq 2000$ K
characteristic of shock excitation. Therefore any objects with observed
ratios of H$_{2}$ $v = 1-0$ $S(1)$ to $v = 2-1$ $S(1) \simeq 2$ ($2-1/1-0
\simeq 0.5$) may be exhibiting UV-excited H$_{2}$ emission. However,
there are no objects in our sample with H$_{2}$ $v= 2-1/1-0$ $S(1) >
0.25$, so we do not have any good candidates for purely UV excitation in
a dust-free environment as modeled by \citet{BD87}.

The presence of dust grains can significantly alter these ratios and
complicate their interpretation. \citet{NATNM07} have computed the
expected NIR H$_{2}$ line fluxes for conditions in circumstellar disks
around young stars, modeling X-ray and UV heating in the presence of
both gas and dust and accounting for thermalization at high gas
densities. They find that the H$_{2}$ $v = 2-1$ $S(1)$ to $v = 1-0$ $S(1)$
line ratio is greatly impacted by the presence of dust grains of
different sizes (see their Figure 17). For UV excited emission,
\citet{NATNM07} find that this line ratio is $\simeq 0.025$ for a power
law dust grain distribution with a maximum size of 10 $\mu$m - 1 mm,
roughly consistent with that expected for a protostar's circumstellar
disk. They find that the grains must be much larger ($\sim10$ cm or
more) for this ratio to approach the dust-free value of 0.5 computed by
\citet{GD95}. This line ratio is much less sensitive to grain size in
the case of X-ray excitation; \citet{NATNM07} find that this value is
close to the \citet{GD95} value of 0.06 for a maximum grain size of 10
$\mu$m - 10 cm. Thus they find that the H$_{2}$ $v = 2-1$ $S(1)$ to $v =
1-0$ $S(1)$ line ratio differs by only about a factor of 2 for a
circumstellar disk with a power law grain size distribution with a
maximum size 10 $\mu$m - 1 mm. \citet{NATNM07} do not consider
collisional excitation by winds or jets, but collisional excitation may
produce a fairly wide range of gas temperatures and H$_{2}$ line ratios
as discussed previously.

If excited by collisions in shocks, ratios of the $v = 1-0$ $S(1)$ to
$v = 1-0$ $S(0)$ emission lines can be used to estimate ortho:para
ratios of molecular hydrogen and to assess the thermal history of the
emitting gas. The values of ortho:para ratios were modeled for C- and
J-type shocks by \citet{WCPdF00}; see also their summary of previous
work. \citet{KRFLP07} and \citet{HPRB98} showed that the $v = 1-0$
$S(1)$ to $v = 1-0$ $S(0)$ emission line ratio directly yields the
molecular hydrogen ortho:para ratio with little sensitivity to H$_{2}$
rotation temperature. Using Eq. 5 of \citet{KRFLP07} and assuming an
H$_{2}$ rotation temperature of 3500~K, we find that the mean
ortho:para ratio for our object sample is $<o/p> = 3.2 \pm 0.8$.
Hydrogen atom exchanges in shocks set the high temperature limit to o/p
$\leq 3$ \citep[e.g., see][]{WCPdF00}. Thus it appears that not all
objects in our sample have molecular hydrogen emission consistent with
production in shocks since a number of objects have o/p $> 3$.
Unfortunately we were unable to use line ratios to diagnose the nature
of the spatially extended H$_{2}$ emissions seen in many objects (see
Table~\ref{tbl-2} and \S 4.1). This emission was generally much weaker
than the point source emission, and it was not detected significantly
in any NIR H$_{2}$ line except $v = 1-0$ $S(1)$ for any object.

\subsection{Molecular Hydrogen Excitation Mechanisms in Observed Protostars}

Our sample has 5 objects with H$_{2}$ $v = 2-0$ $S(1)$ to $v = 1-0$
$S(1)$ line ratios in the 0.025 - 0.06 range, consistent with UV or X-ray
excitation in the presence of dust. Of these, 04264+2433, 04295+2251,
04365+2535, and IRS 43 have relatively low mean H$_{2}$ $v = 1-0$ $S(1)$
line widths, FWHM $\lesssim 15$ km s$^{-1}$. All but 04264+2433 have been
detected in X-rays, so X-ray or UV excitation may be possible for these
protostars. WL 12 is the other protostar with a line ratio in this range,
and it has a broad FWHM $\simeq 30$ km s$^{-1}$ and is associated with a
molecular outflow \citep{BATC96}. Therefore its H$_{2}$ may be
collisionally excited. However, WL 12 has ortho:para ratios of about 4.5,
larger than the o/p $\leq 3$ limit that can be produced in C-shocks or
J-shocks. The objects 04158+2805, 04181+2654, and WL 6 also have
estimated o/p $\geq 3.5$, indicating non-shock excitation even if their
H$_{2}$ rotation temperature is somewhat higher than the assumed 3500~K.
Therefore the NIR H$_{2}$ emission of these objects may not be excited in
jets. The latter 3 objects also have H$_{2}$ radial velocity offset by
less than 5 km s$^{-1}$ from their photospheric velocities. However,
04158+2805 and 04181+2654 have H$_{2}$ FWHM line widths $\geq$ 20
km~s$^{-1}$, clouding a non-mechanical interpretation of their molecular
hydrogen excitation.

Interestingly, there are several objects in the sample whose observed
velocities {\it and} line fluxes suggest quiescent, non-mechanical origins
for their molecular hydrogen emissions. 04361+2547, WL 6, and IRS 67
all have small H$_{2}$ FWHM line widths, small H$_{2}$ velocity offsets
from photospheric velocities, and small H$_{2}$ emission spatial
extents (See Table 2). Interestingly, at least one measurement of the
$v = 1-0$ $S(1)$ to $v = 2-1$ $S(1)$ line ratios is $\sim$0.07 for each
object, similar to the value of 0.06 computed by \citet{GD95} and
\citet{NATNM07} for X-ray excitation of H$_{2}$. Thus these objects 
appear to be the best candidates for non-mechanical excitation of
their molecular hydrogen emissions.

We conclude this discussion by noting that several of the protostars
have NIR ro-vibrational emission properties consistent with collisional
excitation, and some others appear to be good candidates for X-ray and
/ or UV excitation. This is similar to the results found by
\citet{BMTP08} and \citet{BWSLK08} in their surveys of T Tauri stars.
We also find no individual or set of spectral features that are
inconsistent with previous observations of NIR H$_{2}$ emission in
CTTSs. It appears that there is no single NIR line diagnostic that can
clearly identify the excitation mechanisms of H$_{2}$ in the
circumstellar disks and environments of protostars, and correlations
between diagnostics are not strong (\S 3.5). However, the measures of
emission line morphologies, velocity widths, velocity shifts, and line
ratios can constrain the various emission mechanisms when interpreted
within an appropriate theoretical model.

\section{Summary}

We present new observations of near-infrared H$_{2}$ line emission in
a sample of 18 Class I and flat-spectrum low mass protostars, primarily
in the Tau-Aur and $\rho$ Oph dark clouds. We reach the following
conclusions from analyzing these data:

1. All 17 objects found to have NIR H$_{2}$ ro-vibrational line
emission spatially conincident with their continuum sources in an
earlier epoch were also found to have this emission in this new study,
5 -- 10 years later. There appears to be little temporal variation of
this emission on $\sim1$ day time scales. Ten of the 18 protostars
showed H$_{2}$ $v=1-0$ $S(1)$  line emission that was over $1\arcsec$ in spatial extent in
at least one observation (at least one position angle).

2. Nearly all of the protostars have H$_{2}$ $v=1-0$ $S(1)$ line emission
radial velocities within 10 km s$^{-1}$ of their stellar photospheric
line velocities; only 3 objects have H$_{2}$ velocity offsets greater
than or equal to 10 km s$^{-1}$. This evidence suggests that
collisional excitation in jets is unlikely to be the molecular hydrogen
excitation mechanism in many objects.

\begin{figure}
\includegraphics[ scale=0.8,angle=0]{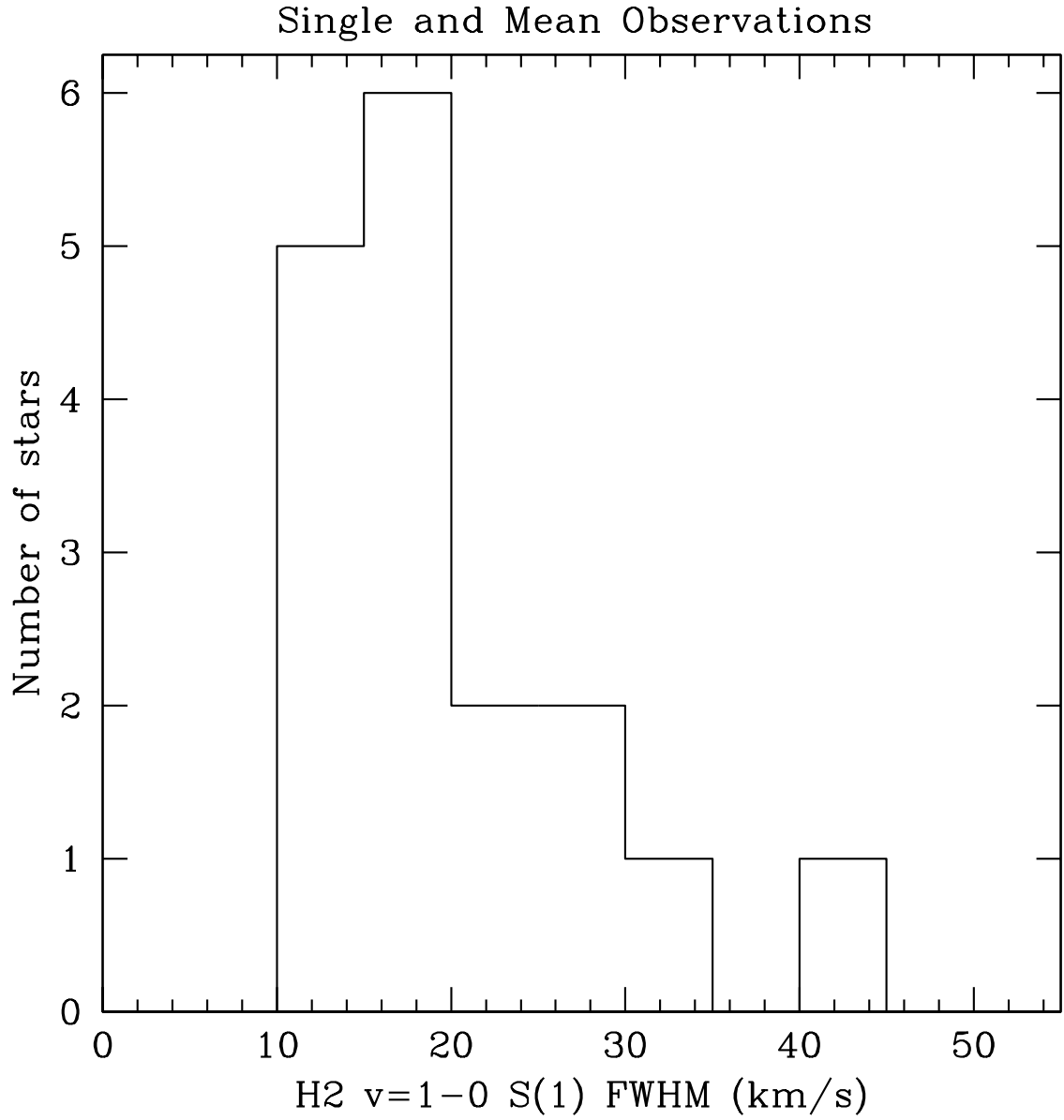}
\caption{Histogram of H$_{2}$ $v = 1-0$ $S(1)$ line FWHM velocities.
The instrumental line width of 17 km s$^{-1}$ has been
subtracted in quadrature from each value before binning. \label{fig3}}
\end{figure}

\begin{figure}
\includegraphics[ scale=0.8,angle=0]{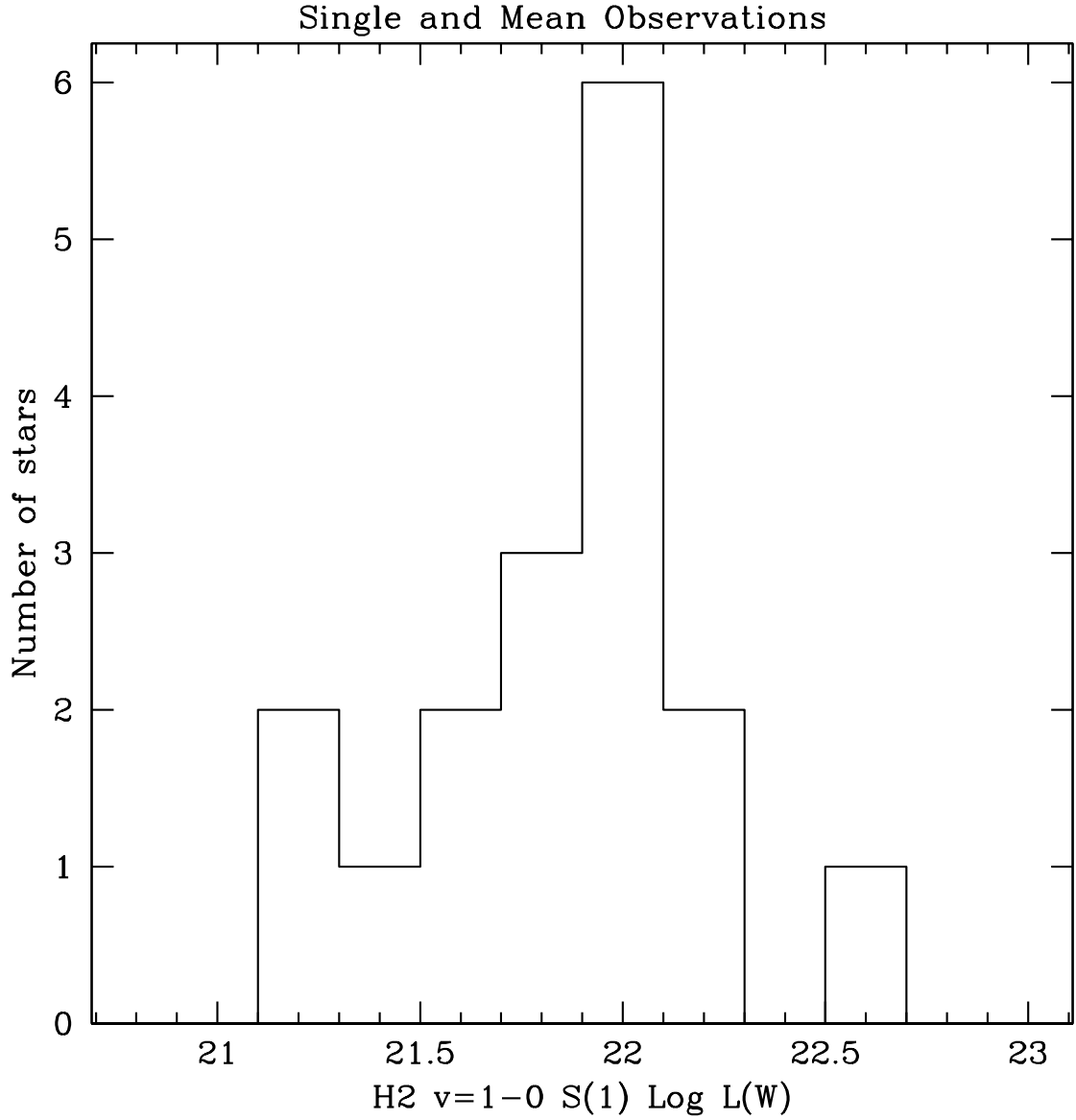}
\caption{Histogram of H$_{2}$ $v = 1-0$ $S(1)$ line luminosities.\label{fig4}}
\end{figure}

\begin{figure}
\includegraphics[ scale=0.8,angle=0]{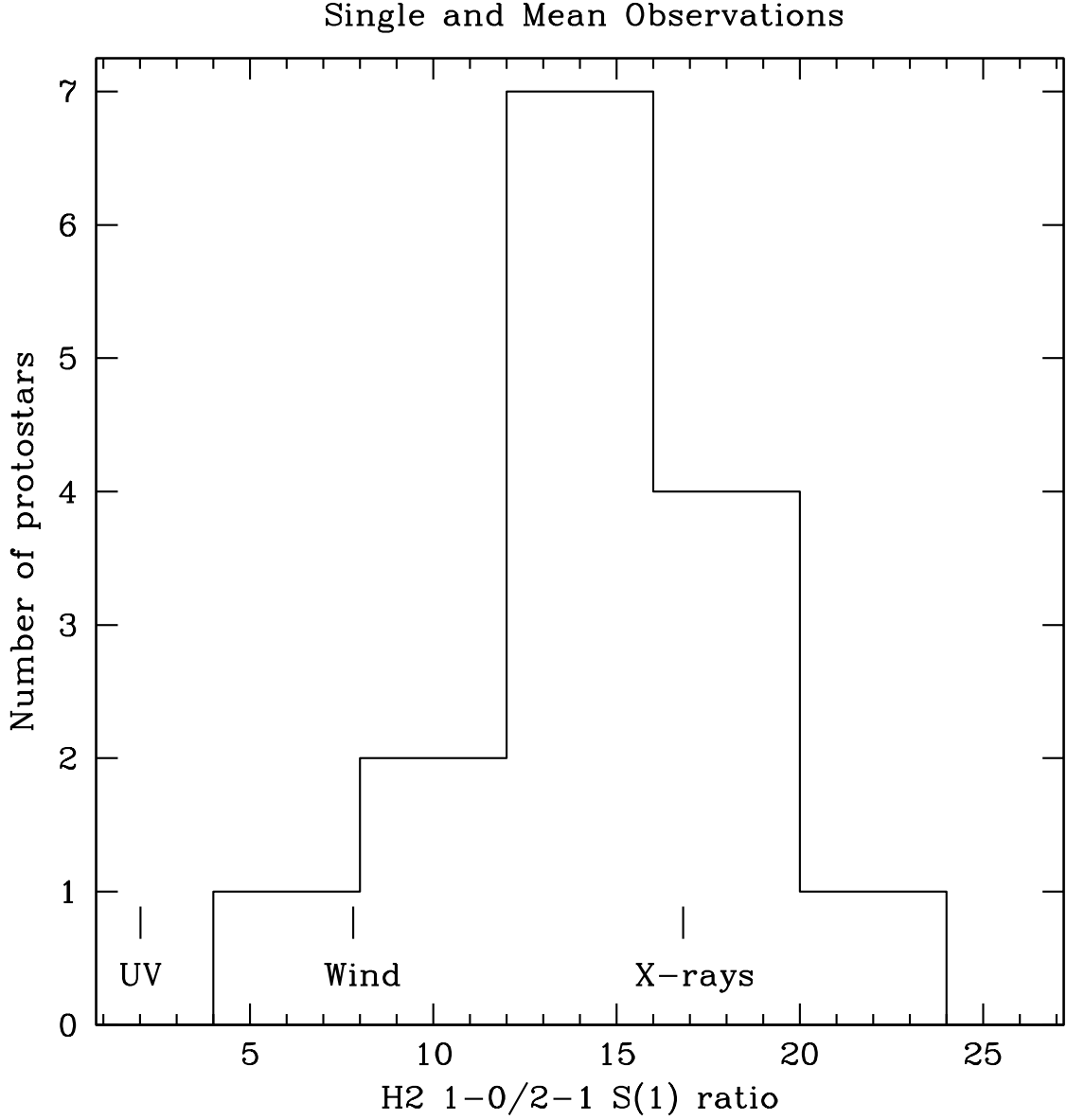}
\caption{Histogram of H$_{2}$ $v = 1-0 / 2-1$ $S(1)$ line ratios.  Values
of 1.9, 7.7, and 17 are indicated, nominally correspond to UV, shock,
and X-ray secondary electron impact excitation respectively in the
presence of no dust \citep{GD95}.  These values are the inverse of
those presented in column 8 of Table 2.
\label{fig5}}
\end{figure}

3. The H$_{2}$ $v = 1-0$ $S(1)$ line FWHM line widths of all 17
protostars with this feature on-source are broader than 10 km s$^{-1}$
(Fig.~\ref{fig3}). Six of the emitting objects (35\%) exhibit FWHM
greater that 20 km s$^{-1}$, and these are good candidates for
collisional excitation in jets or winds. Seven of the emitting objects
(41\%) have point source FWHM line widths below 16 km s$^{-1}$. The
spatially extended H$_{2}$ $v=1-0$ $S(1)$ line emission of two objects
had line widths FWHM $\gtrsim$ 40 km s$^{-1}$, about twice that of
their central point source emission. This is consistent with
collisional excitation by jets or winds.

4. The molecular hydrogen emission line velocities and FWHM values of
at least 3 to 6 of the 17 objects with on-source emission are
consistent with collisional excitation in jets. At least 7 or 8
objects have H$_{2}$ velocity parameters consistent with quiescent
(non-collisional) excitation. There are several objects whose small
emission line widths, small H$_{2}$ -- photospheric radial velocity
differences, and small spatial extents are more consistent with
quiescent molecular hydrogen emission and not collisional excitation.

5. Several of the protostars have H$_{2}$ $v = 2-0$ $S(1)$ to $v =1-0$
$S(1)$ line ratios indicative of X-ray or UV excitation (in the
presence of dust) and are known X-ray emitters. 04361+2547, WL 6, and
IRS 67 are the best examples of such protostars. However, we see no
rapid variation in the H$_{2}$ $\Delta v=1$ $S(1)$ line fluxes on
$\sim$24 hr time scales as might be expected from excitation by X-ray
flaring events.

6. We find that the mean ortho:para ratio for our object sample is
$<o/p> = 3.2 \pm 0.8$. Hydrogen atom exchanges in shocks set the high
temperature limit to o/p $\leq 3$ \citep[e.g., see][]{WCPdF00}. Thus it
appears that not all objects in our sample have molecular hydrogen
emission consistent with production in shocks since a number of objects
have o/p~$> 3$. 04158+2805, 04181+2654, WL 6, and WL 12 are all estimated
to have ortho:para ratios significantly higher than this value. However,
WL 6 is the only protostar with H$_{2}$ line ratios and velocities also
indicative of non-mechanical excitation.

\acknowledgments

We thank D. Hollenbach and U. Gorti for helpful discussions of our data
and its interpretation via theoretical models. We also thank G. Herczeg
for discussing pre-publication data and thank the anonymous referee for
thoughtful suggestions that improved this paper. The Keck Observatory
Observing Assistants H. Hershley and C. Parker are thankfully
acknowledged for assistance with the observations. The authors wish to
recognize and acknowledge the very significant cultural role and
reverence that the summit of Mauna Kea has always had within the
indigenous Hawaiian community. We are most fortunate to have the
opportunity to conduct observations from this mountain. TPG
acknowledges support from NASA's Origins of Solar Systems program via
WBS 811073.02.07.01.89. MB and TPG would like to acknowledge NASA
support via NExScI for travel expenses to the W.M. Keck Observatory for
acquiring the observations for this project.



{\it Facilities:} \facility{Keck (NIRSpec)}, \facility{IRAF}, \facility{2MASS}



\clearpage



\clearpage

\clearpage

\clearpage

\clearpage

\clearpage

\clearpage



\textheight=250mm
\oddsidemargin=-5mm
\topmargin -15mm


\begin{deluxetable}{llrrrrr}
\tablecaption{Journal of Observations \label{tbl-1}}
\tabletypesize{\scriptsize}
\tablewidth{0 pt}
\tablehead{
\colhead{Object} & 
\colhead{Region} & 
\colhead{$\alpha$(J2000)} & 
\colhead{$\delta$(J2000)}  & 
\colhead{UT Date} & 
\colhead{Int. Time} & 
\colhead{Slit PA} \\ 
& 
& 
\colhead{(hh mm ss.s)} & 
\colhead{($\arcdeg$~$\arcmin$~$\arcsec$)}  &
& 
\colhead{(minutes)} & 
\colhead{($\arcdeg$E of N)}
}
 
\startdata
03260+3111B & Per     & 03 29 07.7   & 31 21 58   & 2008 Jan 24  &  20.0  &  103 \\
            &         &              &            & 2009 Jan 25  &  20.0  &  149 \\
03260+3111A & Per     & 03 29 10.7   & 31 21 59   & 2008 Jan 24  &  20.0  &  118 \\
            &         &              &            & 2008 Jan 25  &  12.0  &  180 \\
04108+2803B & Tau-Aur & 04 13 54.9   & 28 11 31   & 2008 Jan 24  &  20.0  & -123 \\
            &         &              &            & 2008 Jan 25  &  25.0  &  121 \\
04158+2805  & Tau-Aur & 04 18 58.2   & 28 12 24   & 2008 Jan 24  &  20.0  &   73 \\
            &         &              &            & 2008 Jan 25  &  20.0  &   63 \\
04181+2654AB  & Tau-Aur & 04 21 11.5   & 27 01 09   & 2008 Jan 24  &  10.0  &   48 \\
            &         &              &            & 2008 Jan 25  &  30.0  &  172 \\
DG~Tau      & Tau-Aur & 04 27 04.8   & 26 06 17   & 2008 Jan 25  &  10.0  &   57 \\
04264+2433  & Tau-Aur & 04 29 30.0   & 24 39 56   & 2009 Jan 24  &  12.0  &   57 \\
            &         &              &            & 2008 Jan 25  &  12.0  & -106 \\
04295+2251  & Tau-Aur & 04 32 32.1   & 22 57 27   & 2009 Jan 24  &  40.0  & -106 \\
            &         &              &            & 2008 Jan 25  &  30.0  &   75 \\
04361+2547  & Tau-Aur & 04 39 13.5   & 25 53 20   & 2008 Jan 24  &  12.0  &   68 \\
            &         &              &            & 2008 Jan 25  &  12.0  & -110 \\           
04365+2535  & Tau-Aur & 04 39 35.2   & 25 41 45   & 2008 Jan 24  &  44.0  &  97 \\
            &         &              &            & 2008 Jan 25  &  33.3  &  99 \\

GSS~30      & Oph     & 16 26 21.4   & -24 23 06  & 2007 Jun 25  &  14.0  & -124 \\
GY~21       & Oph     & 16 26 23.6   & -24 24 38  & 2007 Jun 24  &  20.0  & -56 \\
            &         &              &            & 2007 Jun 25  &  20.0  & -100 \\
WL~12       & Oph     & 16 26 44.1   & -24 34 48  & 2007 Jun 24  &   8.0  & -47 \\
            &         &              &            & 2007 Jun 25  &  10.0  & -114\\
WL~6        & Oph     & 16 27 21.6   & -24 29 51  & 2007 Jun 24  &  20.0  & -72 \\
            &         &              &            & 2007 Jun 25  &  30.0  & -89 \\
IRS~43      & Oph     & 16 27 27.0   & -24 40 50  & 2007 Jun 24  &  45.0  & -78 \\
            &         &              &            & 2007 Jun 25  &  60.0  & -66 \\
YLW~16A     & Oph     & 16 27 27.8   & -24 39 32  & 2007 Jun 24  &  12.0  & -85 \\
            &         &              &            & 2007 Jun 25  &  12.0  & -133 \\
IRS~67      & Oph     & 16 32 01.1   & -24 56 45  & 2007 Jun 24  &  60.0  & -59 \\

SVS~2       & Ser     & 18 29 56.8   &  01 14 46  & 2007 Jun 25  &  28.0  & -85 \\

\\
\enddata

\end{deluxetable}

\clearpage

\begin{deluxetable}{llrrrrrrrrr}
\rotate
\tablewidth{0pc}

\tablecaption{Protostar H$_{2}$ Line Analysis\label{tbl-2}}
\tablehead{
\colhead{Source}      & 
\colhead{UT Date}  &
\colhead{1--0 S(0)} & 
\colhead{1--0 S(1)} & \colhead{2--1 S(1)} &
\colhead{A$_v$\tablenotemark{b}} & \colhead{1--0 S(1)} & 
\colhead{2--1/1--0} &
\colhead{FWHM\tablenotemark{c}} & \colhead {V(H$_{2}$ - *)\tablenotemark{d}} &
\colhead{H$_{2}$ 1--0 S(1)}\\ [0.2 ex]
\colhead{}            & 
\colhead{}            & 
\colhead{EW (\AA)\tablenotemark{a}} & 
\colhead{EW (\AA)\tablenotemark{a}} & \colhead{EW (\AA)\tablenotemark{a}} &
\colhead{(mag)} & \colhead{Log L(W)} & 
\colhead{S(1)}  & 
\colhead{(km s$^{-1}$)} & \colhead{(km s$^{-1}$)} &
\colhead{extent (\arcsec)}
}
\tablecolumns{11}
\startdata
03260+3111B & 2008 Jan 24\tablenotemark{e} & 0.50 & 2.12 & 0.49 & 3.5 & 22.0 & 0.25 & 43
& \ldots & $\gtrsim 10$\\
            & 2008 Jan 25 & 0.51 & 2.69 & 0.39 & 3.5 & 22.1 & 0.14 & 44 
            & \ldots & $\lesssim 1$ \\
\\
03260+3111A\tablenotemark{f} & 2008 Jan 24 & $<0.1$ & $<0.1$ & $<0.1$ & \ldots & \ldots &
\ldots & \ldots & \ldots & $\sim 7$ \\
            & 2008 Jan 25 & $<0.1$  & $<0.1$  & $<0.1$  & \ldots & \ldots &
            \ldots & \ldots & \ldots & $\gtrsim 10$ \\
\\
04108+2803B & 2008 Jan 24 & 0.92 & 4.29 & 0.39  & 19 & 22.0  & 0.08 & 18
            & -1 & $\sim 1$\\
            & 2008 Jan 25 & 0.69 & 3.43 & 0.29 & 19 & 21.9 & 0.07 & 18
	    & -1 &  $\sim 1$\\
\\
04158+2805  & 2008 Jan 24 & 0.49 & 2.57 & 0.26: & 3.1 & 21.1 & 0.08: &
            20 & -4 & $< 1$\\ 
            & 2008 Jan 25 & 0.35 & 2.70 & 0.22: & 3.1 & 21.2 & 0.06: &
	    20 & -4 & $\sim 2$\\
\\
04181+2654AB & 2008 Jan 24 & 0.23 & 1.46 & 0.04:: & 23 & 22.0 &
             \ldots & 24 & -1 & $\sim 1$\\
	     & 2008 Jan 25 & 0.39 & 1.37 & 0.00:: & 23 & 21.9 & \dots
	     & 25 & 8 & $< 1$\\
\\
DG Tau      & 2008 Jan 25 & 0.14 & 0.54 & 0.06:: & 0 & 22.0 &
            0.08:: & 14 & -9 & $< 1$\\
\\
04264+2433  & 2008 Jan 24 & 1.56 & 5.78 & 0.53 & 6.1 & 21.7 & 0.06 & 10
            & 4 & $\sim 2$\\
	    & 2008 Jan 25 & 1.44 & 6.09 & 0.52 & 6.1 & 21.7 & 0.06 & 13
	    & -4 & 6\\
\\
04295+2251  & 2008 Jan 24 & 0.59 & 2.42 & 0.14: & 17 & 22.0 & 0.04: & 
            16 & \ldots & $\sim 2$\\
            & 2008 Jan 25 & 0.59 & 2.43 & 0.16: & 17 & 22.0 & 0.05: &
	    16 & \ldots & $\sim 2$\\
\\
04361+2547  & 2008 Jan 24 & 1.06 & 4.11 & 0.45 & 22 & 22.2 & 0.08 & 14
            & \ldots & 0\\
	    & 2008 Jan 25 & 0.66 & 2.10 & 0.21: & 22 & 21.9 & 0.07: & 
	    13 & \ldots & 0\\
\\
04365+2535  & 2008 Jan 24 & 0.57 & 2.58 & 0.18: & 19 & 21.9 & 0.06: &
            19 & \dots & 0\\
            & 2008 Jan 25 & 0.35 & 1.59 & 0.09: & 19 & 21.6 & 0.05: &
	    16 & \dots & 0\\
\\

GSS 30    & 2007 Jun 25 & 0.53 & 2.35 & 0.20 & 20 & 22.6  & 0.07 & 
          25 & -11 & 2\\
\\
GY 21\tablenotemark{g}     & 2007 Jun 24 & 0.37 & 1.80 & 0.20 & 16 & 21.8 & 0.07 &  
          28 & -11 & $\sim 1$\\
          & 2007 Jun 25 & 0.52 & 2.21 & 0.20 & 16 & 21.9 & 0.07 &  
	  29 & -10: & $\sim 1$\\
\\
WL 12     & 2007 Jun 24 & 1.43 & 12.4 & 0.79 & 18 & 22.1 & 0.06  & 
          32 & -2 & $\sim 2$\\
          & 2007 Jun 25 & 2.06 & 11.0 & 0.73 & 18 & 22.1 & 0.06  & 
	  29 & -12 & $\sim 1$\\
\\
WL 6      & 2007 Jun 24 & 0.06: & 0.34 & 0.06: & 37 & 21.6 & 0.15:   
          & 15 & 0 & 0\\
          & 2007 Jun 25 & 0.09 & 0.37 & 0.03: & 37 & 21.6 & 0.07: 
	  & 16 & -3 & 0\\
\\
IRS 43\tablenotemark{h}    & 2007 Jun 24 & 0.09 & 0.28 & 0.02:: & 33 & 21.8 & 0.05:: 
          & 9 & 0 & $\sim 3$\\
          & 2007 Jun 25 & 0.17 & 0.64 & 0.04: & 33 & 22.1 & 0.05:: 
	  & 22 & -4 & $\sim 3$\\
\\
YLW 16A   & 2007 Jun 24 & 0.32 & 1.41 & 0.20 & 17 & 21.6 &  0.12  & 25 & 
	  0 & $\sim 2$\\
          & 2007 Jun 25 & 0.31 & 1.25 & 0.13 & 17 & 21.6 & 0.09 
	  & 28 &  -4 & $\sim 3$\\
\\	   
IRS 67    & 2007 Jun 24 & 0.13 & 0.37 & 0.03:: & 22 & 21.1 & 0.07:: & 
          15 & 0 & $\sim 1$\\

\\
SVS 2\tablenotemark{i}     & 2007 Jun 25 & 0.09 & 0.39 & 0.01::: & 0 & 21.5 & \ldots
          & 13 & 19 & $\sim 3$\\
\enddata

\tablenotetext{a}{Positive equivalent widths indicate emission.
Uncertainties are $\sim 0.02 \AA$ for 2007 data and $\sim 0.05 \AA$ for
2008 data.}

\tablenotetext{b}{V magnitude extinction was computed using each objects 2MASS
$JHK$ colors, an extinction law, and estimated intrinsic CTTS locus colors as 
explained in the text in \S 3.4}

\tablenotetext{c}{ The mean FWHM velocity of the 1--0 S(1) H$_{2}$
line, where the intrinsic instrumental line width of 17 km s$^{-1}$
has been removed in quadrature.}

\tablenotetext{d}{ The radial velocity of the stellar photosphere
subtracted from the mean radial velocity of the 1--0 S(0), 1--0 S(1),
and 2--1 S(1) H$_{2}$ emission lines (uncertain lines not used). No
data indicate that the star lacked either H$_{2}$ or photospheric lines.}

\tablenotetext{e}{ Spectrum displays $\Delta v = 2-0$ CO emission.}

\tablenotetext{f}{Extended H$_{2}$ emission is spatially displaced
from the star.  No H$_{2}$ emission is coincident with the stellar
continuum (as also observed by Doppmann et al.  2005).}

\tablenotetext{g}{The H$_{2}$ emission slightly spatially displaced
from the stellar continuum has velocity FWHM $\Delta v \sim 60$ km
s$^{-1}$, about twice that of the H$_{2}$ emission spatially
coincident with the stellar continuum.}

\tablenotetext{h}{Extended H$_{2}$ emission shows velocity structure
with a maximum FWHM $\Delta v \sim 40$ km s$^{-1}$, about twice that
of the H$_{2}$ emission spatially coincident with the stellar
continuum.}

\tablenotetext{i}{The extended H$_{2}$ emission is separated from the 
stellar continuum and its line emission by $\sim 2\arcsec$. The
extended emission has FWHM $\Delta v \sim 24$ km s$^{-1}$, about twice that
of the H$_{2}$ emission spatially coincident with the stellar
continuum. H$_{2}$ line strength ratio was not calculated due to the
very low value and high uncertainty ($\sim50$\%) of the 2--1 S(1) line 
measurement.}

\end{deluxetable}



\end{document}